# Augmenting Learning with Augmented Reality: Exploring the Affordances of AR in Supporting Mastery of Complex Psychomotor Tasks

Dong Woo Yoo, Northeastern University, yoo.d@northeastern.edu
Sakib Reza, Northeastern University, reza.s@northeastern.edu
Nicholas Wilson, Harvard University, nicholascwilson@gmail.com
Kemi Jona, Northeastern University, k.jona@northeastern.edu
Mohsen Moghaddam, Northeastern University, mohsen@northeastern.edu

**Abstract:** This research seeks to explore how Augmented Reality (AR) can support learning psychomotor tasks that involve complex manipulation and reasoning processes. The AR prototype was created using Unity and used on HoloLens 2 headsets. Here, we explore the potential of AR as a training or assistive tool for spatial tasks and the need for intelligent mechanisms to enable adaptive and personalized interactions between learners and AR. The paper discusses how integrating AR with Artificial Intelligence (AI) can adaptably scaffold the learning of complex tasks to accelerate the development of expertise in psychomotor domains.

## Introduction

Augmented Reality (AR) technologies are increasingly being used to create engaging learning experiences that provide learners the ability to engage in real-world activities while supported with contextualized multimedia scaffolds. In this paper, we present a prototype AR tutoring system that aims at fostering the learning and adaptability of learners in industrial workplaces through just-in-time delivery of instructions superimposed on the physical workspace. The context of the prototype is related to precision inspection work in the manufacturing industry (e.g., aerospace, medical), where complex parts must be produced with extreme precision and continuously inspected using a wide range of special-purpose gauges under exacting specifications. We chose this area because it requires learners to master complex "psychomotor-intensive" tasks that can serve as analogues to many other such application areas and the affordances of AR uniquely lend themselves to physical tasks being done in the real world. At the macro-level, the country faces a growing manufacturing skills gap that is driven by the need for adaptive expertise in various industries ranging from manufacturing to healthcare and defense (Brown & Wilson, 2019). If successful, our solution may help address this pressing workforce need and potentially empower workers with physical or cognitive disabilities.

Our AR prototype provides users with an interactive experience where they can intuitively navigate through inspection instructions using hand gestures and natural dialogue, receive personalized data-driven feedback, and access the information they need at any moment in the form of textual instructions, potential consequences of errors, and immersive 3D animations. In our demo, described in the next section, users will be guided in completing these tasks with AR guidance that can adapt to their level of expertise and needs for instructions or feedback during training. The system is being designed to help users complete such tasks faster, more independently, and with fewer errors over time. We will also investigate the extent to which users can then transfer the learned skills between different tasks that involve precision instruments and multiple parts. The future version of the AR system that is currently in development (see **Future Directions**) will incorporate advanced AI capabilities in computer vision, natural language processing, and multimodal data fusion to support activity understanding (e.g., action recognition, error detection), user modeling (e.g., cognitive load detection), and rule-based inferences to further tailor the instructions to the individual needs of users in real time.

## Technological Setup and Demonstration

The proposed demonstration aims at showcasing novel human-machine teaming capabilities through AR that have the ability to support the learning of "spatial tasks" on the job, in an adaptive, real-time, and personalized way. Examples of such psychomotor-intensive tasks range from controlling an automated production cell in a factory to operating advanced medical equipment in a hospital. Our vision is to build a new category of intelligent tutoring systems that supports complex psychomotor tasks that involve interactions between humans and tools and equipment (e.g., a robot, an fMRI scanner, a CNC machine, a metal 3D printer) in authentic settings through AR.

Without loss of generality, we have adopted a simple yet practical use case in precision inspection to test and validate the affordances of the proposed AR technologies in controlled laboratory settings before scaling it up to more complex and practical scenarios. This demo will be centered on the precision inspection use case, which is a critical step in many advanced manufacturing processes in sectors such as aerospace, medical, and defense. This is the recurrent process of examining minor, major, or critical characteristics of the part or assembly using a variety of standard or unique-purpose gauges. An "expert" in this context has an in-depth knowledge of materials and processes, and the necessary skills to properly use inspection devices (e.g., setup, measurement) and interpret product manufacturing information (e.g., drawings, characteristics, tolerances). Precision inspection processes often require a thorough understanding of complex 2D CAD blueprints and job sheets, precise measurement of dozens of characteristics, and meticulous adjustment of process parameters (e.g., machine offsets) accordingly, considering various material and tool properties. To better comprehend the complex nature of this task, consider a jet engine as an example. A jet engine comprises several static and rotating parts that must be machined and assembled to perfection with tolerances ranging from 0.005" to 0.0001" (human hair is approximately 0.001" thick). Even slightly out-of-tolerance parts can lead to catastrophic failure of the engine. To perform proper inspection during production, the operator must meticulously stage and calibrate the gauge on the master and on the part, then remove the master, retract the dials, and "feel" that the gauge navigates over the characteristic being measured.

This interactive demo will show how our novel AR technology can augment the ability of learners to gain and transfer the skills necessary to carry out psychomotor intensive tasks by supporting adaptive provision of learning instructions and feedback through immersive animations, voice commands, audio instructions, and flexible navigation through the training system. In the precision inspection use case, for example, the intelligent AR system enables personalized and on-demand visualization of part/gauge setups, calibration techniques, how and when to retract the gauge to allow it to pass over the interior obstacles that the learner cannot see, and the motion/feel necessary to obtain accurate readings. This demo will also show how AR can visually put the learner in touch with the other dimensions that need to be checked. Further, if the parts are out-of-tolerance or non-conformant, AR can recommend diagnostics and actions, and inform the learner of the potential consequences to facilitate spatial and causal reasoning and instill the importance of critical measurements. The demonstration setup will consist of a set of precision inspection gauges that the participants can use to inspect various characteristics of a number of 3D printed parts following the guidance provided by the AR system through Microsoft HoloLens 2 headsets (Figure 1). The AR tutoring system allows for the synchronizing of training instructions with the physical equipment and environment through text, 3D animations, videos, and audio cues. It provides real-time guidance to users during hands-on-training and task performance to ensure proficient skills. The AR app is designed for use with Microsoft HoloLens 2; however, it is compatible with a range of AR and headsets (e.g., Magic Leap 2, Varjo XR-3) and mobile devices such as tablets and smartphones.

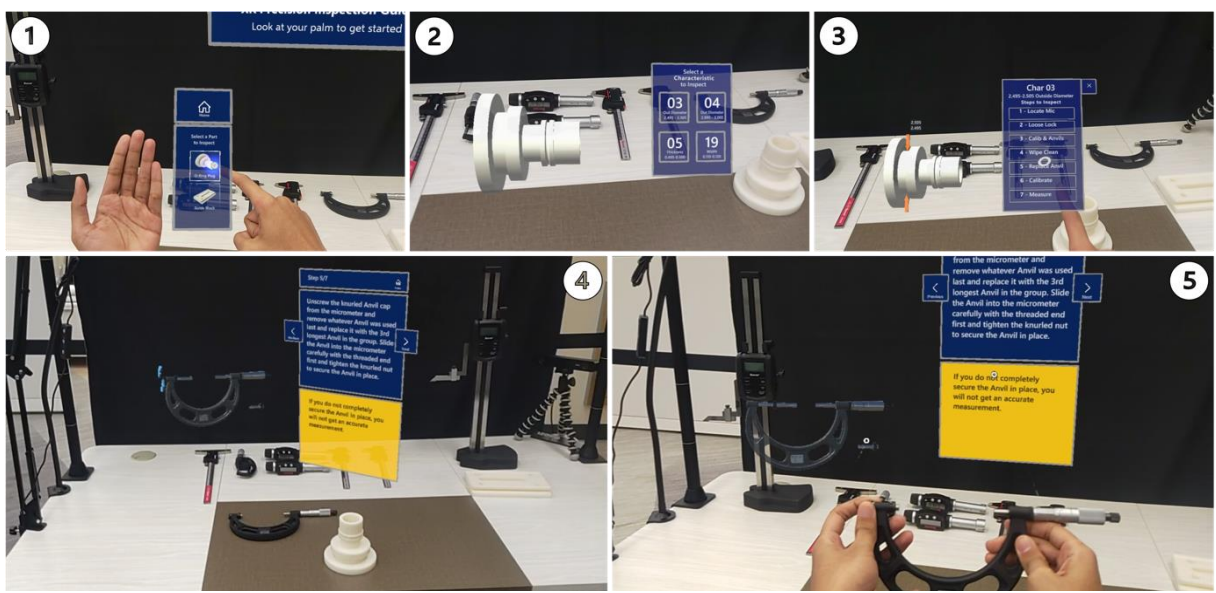

**Figure 1.** *Overview of the AR tutoring system. (1) Part selection: Upon opening the AR app, the user can select the part to be inspected from a hand menu. (2) 3D characteristics menu: Once the part to be inspected is chosen, a 3D model of that part will be displayed with a menu with all the characteristics. When the user hovers her finger*

above a particular characteristic in the menu, the characteristic will be highlighted in 3D, following the industry standards for Geometric Dimensioning and Tolerancing (GD&T). (3) Flexible navigation: When the user chooses a characteristic to inspect, a coarse-level step list is first displayed comprising a short description of each step at once. This would allow the user to directly navigate to a desired step without having to go through the fixed procedure of the task. (4-5) Multimedia instructions: Instructions are provided in different mediums like immersive animation, text, and audio. Animations are designed to be self-explanatory, but to give the user more details we give them text instructions with the audio narration of the same text.

## Lessons from Prior Work

In a previous study, we conducted a study to examine the potential of AR to facilitate the learning of a marine engine sub-assembly task. Generally, experienced assemblers use standard hand/power tools to assemble the fuel cell following instructions on a one-page sheet with technical drawings and materials list. The main challenge for the manufacturer is training novice workers, often with a background in mechanics or machinists, which can take several weeks or months and is often done by experienced workers which is not desirable. The core objective of the study was to investigate if and how AR can help address this challenge (Figure 2).

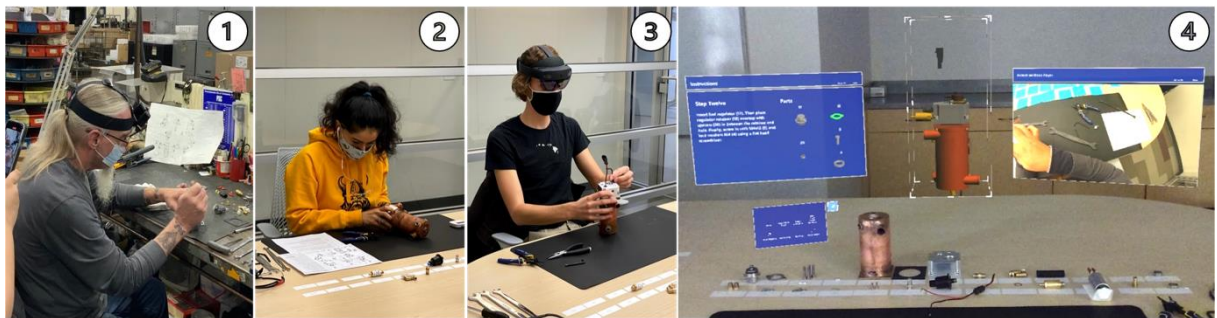

**Figure 2.** *(a) Data collection and expert capture. (b) Paper group. (c) AR group. (a) The AR app.*

*Apparatus*: The experiments involved the assembly of a fuel module using standard mechanical tools. The AR app was developed using Unity and Mixed Reality Toolkit, and includes expert capture GoPro videos with vocal cues, textual descriptions and images of parts, and interactive 3D CAD animations. *Participants*: 20 Mechanical and Industrial Engineering students were recruited as participants (6 females, 11 undergrads/9 grads, 2 URMs) and divided into two groups: AR and paper. A questionnaire was used to collect their demographics and related prior experiences to counterbalance the groups. All participants received initial online training on the AR app and hardware (HoloLens 2). *Procedure*: Each participant performed three assembly cycles in separate dates using their designated mode of instruction and returned after a few days to perform a final assembly using the opposite mode of instruction. At the end of each session: (1) the experimenters recorded time to completion, number of errors, frequency of help-seeking behavior, and the types of errors and questions, and (2) the participants reported their cognitive load (NASA-TLX), self-efficacy, experience with HoloLens/AR app, and general feedback through structured and open-ended questions. *Hypotheses*: AR significantly improves time-to-completion, independence, cognitive load, and task competency compared to paper instruction. *Findings*: AR reduces the number of errors by 31-84%. The task completion times of the two groups are about the same; however, that was partly due to the unfamiliarity of participants with AR and some technical issues. Further, most participants reported absolute independence from AR after two/three cycles, which points to the effectiveness of AR in improving task competency, and yet its low utility as an "assistive tool" for routine tasks. Further, several participants suggested devising interactive help and voice command systems (Moghaddam et al., 2021).

## Future Directions

We envision a long-term research agenda to identify the specific affordances of AR that facilitate the progression of novice-to-expert development of skill acquisition in complex psychomotor-intensive tasks (with an initial focus on precision manufacturing tasks). Complex precision manufacturing tasks rely heavily on the development of spatial reasoning abilities (i.e., visualizing how parts interact or parts of machines operate). To that end, our work investigates the impacts of AR not only on user performance, but how users reason their way through psychomotor tasks that involve multiple parts and measurement tools (gauges) on their way to developing expertise.

A significant amount of research in the field of learning sciences has carefully examined how novices and experts across many disciplines from chess players to physicians use different ontologies (Chi & VanLehn, 2012) when creating solutions to problems involving complex systems (Jacobson, 2021; Jacobson & Wilensky, 2006). Hmelo-Silver and Pfeffer (2004) argued that novices tend to use simplified, rule-based methods and heuristics that prefer predictable outcomes and explain phenomena in terms of individual components when solving problems whereas experts are more likely to approach complex problems as nonreductive and dynamic, particularly with regards to identifying the causal mechanisms of phenomena. Moreover, novices are likely to focus on surface or structural features of a system almost exclusively in problem solving tasks, on the other hand, experts attend to these features in addition to the mechanical behaviors and functional characteristics of complex system' components, displaying more nuanced and refined spatial reasoning processes.

The data and analyses collected during the AR application will also allow us to explore data-driven approaches to derive knowledge that AR could facilitate human-machine teaming (Figure 3). We will evaluate the use of data-driven user modeling methods to model user behavior, specifically to identify and address errors during training by comparing user performance to that of an expert. We will investigate various techniques for this modeling process and begin by attempting to create these models through automated analysis of task behavior and cluster inferred models of performance into categories (Sequeira & Marsella, 2018). The National Strategy for Advanced Manufacturing (2022) presents a comprehensive strategy to address workforce development and education. The strategy also emphasizes the importance of workforce development and education, including K-12 and higher education, in preparing workers for the advanced manufacturing employment of the future. This AR application aligns with the national priorities for addressing the growing skills mismatch and the changing nature of skill requirements with the adoption of new technologies, especially in the aftermath of COVID-19.

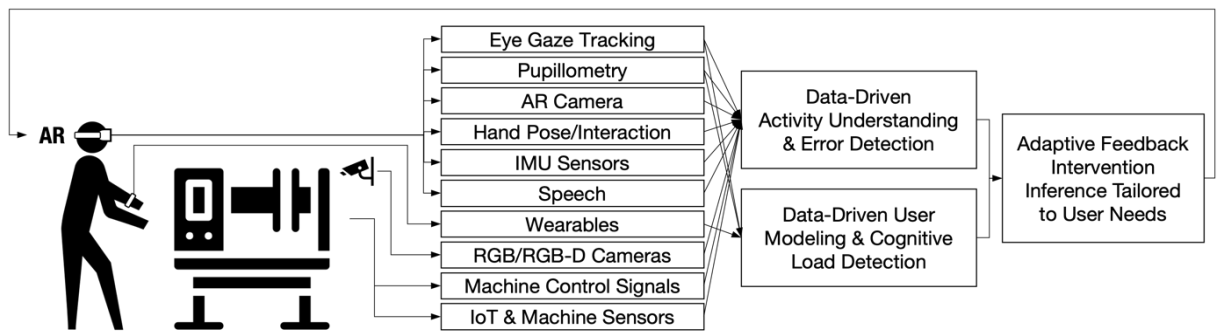

**Figure 3.** *Vision of intelligent AR tutoring system enabled by multimodal sensory data to support activity understanding and user modeling for adaptive and personalized training feedback.*

## Acknowledgments

This material is based upon work supported by the National Science Foundation (NSF) under the Future of Work at the Human-Technology Frontier Grant No. 2128743. Any opinions, findings, and conclusions or recommendations expressed in this material are those of the author(s) and do not necessarily reflect the views of the NSF. We thank Robert Roy, Akhil Ajikumar, Parisa Ghanad Torshizi, and Hamid Tarashiyoun.